\documentclass[runningheads,10pt]{llncs}
\usepackage{graphicx}
\usepackage{amsmath,amsfonts}
\usepackage{soul}
\usepackage{color}
\usepackage{hyperref}

\newtheorem{algorithm}{Algorithm}
\begin{document}
%
\title{Factorizing time evolution into elementary steps
}
%
%
\author{David Edward Bruschi\inst{1}\orcidID{0000-0002-3816-5439}}
\authorrunning{D. E. Bruschi et al.}
%
\institute{Institute for Quantum Computing Analytics (PGI-12), Forschungszentrum J\"ulich, 52425 J\"ulich, Germany\\
\email{david.edward.bruschi@posteo.net}}
\maketitle              
\begin{abstract}
We propose an approach to factorize the time-evolution operator of a quantum system through a (finite) sequence of elementary operations that are time-ordered. Our proposal borrows from previous approaches based on Lie algebra techniques and other factorization procedures, and requires a set of optimization operations that provide the final result. Concretely, the algorithm produces at each step three optimal quantities, namely the optimal duration of the desired unitary operation, the optimal functional dependence of the driving function on the optimal time, and the optimal elementary Hermitian operation that induces the additional unitary operation to be implemented. The resulting sequence of unitary operations that is obtained this way is sequential with time. We compare our proposal with existing approaches, and highlight which key assumptions can be relaxed for practical implementations. 

\end{abstract}
%
%
\section*{Introduction}
Studying the time evolution of a system is paramount in physics. While the evolution of classical systems can be, in principle, known with arbitrary precision, quantum mechanical systems are inevitably affected by the uncertainty that is inherent in quantum mechanics. To study a closed quantum system one must first determine its initial state, and subsequently evolve it in time according to the von Neumann equation \cite{Messiah:1961}. The evolution is implemented by a unitary operator that is induced by the \textit{Hamiltonian} of the system, which contains all of the relevant information that characterize the interaction between the system's constituents. If the application of the time-evolution operator on the initial state can be obtained explicitly, in principle it is possible to fully characterize the system by computing any quantity of interest at any time. However, it rarely occurs that such action is known. In practice, physical systems are composed by extremely large numbers of constituents and, even in the cases where they can be modelled as a few interacting quantum degrees of freedom, the exact action of the time-evolution operator on an arbitrary state is extremely hard to obtain.

Given a Hamiltonian defined by a linear combination with real (functional) coefficients of many independent Hermitian operators, two natural questions can be posed: \textit{are there ways to approximate the time-evolution operator faithfully at any time $T$}? Furthermore, \textit{what conditions on the Hamiltonian allow to factorize a time evolution operator into a (finite) product of known elementary operations?}

These two questions have been investigated for some time. The first one has been approached through a variety of different techniques aimed at obtaining reliable approximations \cite{Leforestier:Bisseling:1991,Kosloff:1994,Tannor:2008,Schaefer:Tal-Ezer:2017}. The second question has been tackled by employing tools from Lie algebra and noting that the operators that define the Hamiltonian, together with their commutation relations, can be lifted to abstract Lie group theory \cite{Wei:Norman:1963,Wei:Norman:1964,Barenco:Bennett:1995}. Similar theoretical work also studied the way to compare (unitary) operators \cite{Halmos:1972,Ando:Sekiguchi:1973,Aiken:Erdos:1980}. Finally, a recent revival of interest has brought novel insights \cite{RauschdeTraubenberg:Slupinski:2006}. Implementations using symplectic geometry \cite{Bruschi:Lee:2013}\footnote{For a less explicit work in this direction see \cite{Brown:Martin-Martinez:2013}.}, as well as solutions obtained for specific classes of systems \cite{Bruschi:Xuereb:2017,Bruschi:2019,Bruschi:Paraoanu:2019,Bruschi:2020} are now available.
Nevertheless, only recently a general result on which classes of Hamiltonians can be factorized into a product of finite terms has been obtained \cite{Bruschi:Zeier:2020}.

In this work we propose a different approach to factorizing the time evolution operator into a product of individual operations obtained following a Lie-algebra approach that are sequential in time. This decomposition is obtained through an optimization algorithm and therefore applies only for a time \text{fixed} a priori. In case one wishes to factorize the time-evolution operator that is applied up to a different time, the optimization algorithm must be repeated from the start. 

\section{Factorizing time evolution}
In this section we set up the problem and introduce the necessary tools. A detailed introduction to the formalism is left to the literature \cite{Wei:Norman:1963,Wei:Norman:1964,Barenco:Bennett:1995,RauschdeTraubenberg:Slupinski:2006,Bruschi:Lee:2013}. 

\subsection{Factorized time-evolution operators}
The information of a quantum system is encoded at time $t$ in its state $|\psi(t)\rangle$. The normalized (pure) state $|\psi(t)\rangle$ satisfies the Schr\"odinger equation $\frac{d}{dt}|\psi(t)\rangle=-\frac{i}{\hbar}\,\hat{H}(t)|\psi(t)\rangle$, where $\hat{H}(t)$ is the \textit{Hamiltonian} of the system (which can be time-dependent). More importantly, the Schr\"odinger equation can be (formally) solved: the solution reads $|\psi(t)\rangle=\hat{U}(t)\,|\psi(0)\rangle$, where $|\psi(0)\rangle$ is the initial state of the system, and $\hat{U}(t)$ is the \textit{time-evolution} operator defined by
\begin{align}\label{time:evolution:operator}
\hat{U}(t):=\overset{\leftarrow}{\mathcal{T}}\exp\left[-\frac{i}{\hbar}\int_0^{+\infty}dt'\,\hat{H}(t')\right],
\end{align}
where $\overset{\leftarrow}{\mathcal{T}}$ is the time-ordering (from right to left) operator. Note that when the Hamiltonian is constant, equation (\ref{time:evolution:operator}) simply reduces to $\hat{U}(t)=\exp\bigl[-\frac{i}{\hbar}\,\hat{H}\,t\bigr]$.

The Hamiltonian $\hat{H}(t)$ is a Hermitian operator and, in general, can be expanded as $\hat{H}(t)=\sum_{n\in\mathcal{I}}h_n(t)\,\hat{G}_n$, where $h_n(t)$ are real time-dependent functions, $\hat{G}_n$ are time-independent Hermitian operators and $\mathcal{I}$ is an appropriate set of labels. As has been already noted, the expression $\hat{H}(t)=\sum_{n\in\mathcal{I}}h_n(t)\,\hat{G}_n$ might contain many terms, therefore making the action of (\ref{time:evolution:operator})  on a quantum state difficult to implement concretely. 
Let us now define the following:
\begin{definition}[Factorized operator] \label{Definition:one}
Given a Hamiltonian $\hat{H}(t)$ with expression $\hat{H}(t)=\sum_{n\in\mathcal{I}}h_n(t)\,\hat{G}_n$, a set of indices $\mathcal{J}\supseteq\mathcal{I}$, and a set of real, time-dependent functions $F_p(t)$ with $p\in\mathcal{J}$, we say that the time-evolution operator $\hat{U}(t)$ is \emph{factorized} if it can be put in the form
\begin{align}\label{time:evolution:operator:decoupled:old}
\hat{U}(t)=\prod_{p\in\mathcal{J}}\exp\left[-i\,F_p(t)\,\hat{G}_p\right]
\end{align}
for appropriate $\{\hat{G}_p\}|_{p\in\mathcal{J}}$, and the coefficients $F_p(t)$ are functions of the coefficients $\{h_n(t)\}$ through a set of ordinary, nonlinear, coupled differential equations. The set $\mathcal{J}$ need not be finite even if $\mathcal{I}$ is.
\end{definition}
Given Definition~\ref{Definition:one}, we ask the following important question: \textit{under which conditions a time-evolution operator \emph{can} be factorized}?
The standard way to attempt a factorization of the form (\ref{time:evolution:operator:decoupled:old}) is to use the well-known Baker-Campbell-Hausdorff formula \cite{Wei:Norman:1964}, which in one of its incarnations reads $\exp[\hat{A}]\exp[\hat{B}]=\exp[\hat{Z}]$ where $\hat{Z}=\hat{A}+\hat{B}+\frac{1}{2}[\hat{A},\hat{B}]+\frac{1}{12}[\hat{A},[\hat{A},\hat{B}]]-\frac{1}{12}[\hat{B},[\hat{A},\hat{B}]]+\ldots$ Clearly $\hat{Z}=\hat{A}+\hat{B}$ if $\hat{A}$ and $\hat{B}$ commute.
However, this formula requires in general a brute-force calculation of (potentially infinite) commutators, which demonstrates the difficulty of the process and is exacerbated further by the time-ordering. 

\subsection{Hamiltonian Lie algebra}
Regardless of the apparently daunting perspective of factorizing the time-evolution operator, in the past decades an increasing number of studies has tackled the problem using mathematical tools stemming from Lie algebra theory. They have addressed both the general theoretical aspects \cite{Wei:Norman:1963,Wei:Norman:1964,Bruschi:Lee:2013}, as well as the concrete model-dependent ones \cite{Bruschi:Xuereb:2017,Bruschi:2019,Bruschi:Paraoanu:2019,Bruschi:2020}. So far, it has been unclear what are the general assumptions that guarantee that a factorization can be obtained, and when such factorization is finite. More precisely, in those cases where $\hat{U}(t)$ admits a factorization (\ref{time:evolution:operator:decoupled:old}), it is important to know what are the conditions for $|\mathcal{J}|<\infty$. Recent work has investigated exactly this aspect, and has found the only classes of Hamiltonians where the factorization is finite \cite{Bruschi:Zeier:2020}. Note that, in such case, the Baker-Campbell-Hausdorff formula would also give a finite expression. It would not, however, inform us under which conditions such finite expression occurs.

In general, most of these studies are based on the fact that the Hamiltonian Lie algebra induced by the \textit{time-independent} operators $\hat{G}_n$ used to decompose the Hamiltonian is finite. Concretely, the Lie algebra $\mathfrak{g}$ is generated by $\{\hat{G}_n\}|_{n\in\mathcal{I}}$ and is closed with respect to all commutators $[\hat{G}_n,\hat{G}_m]$, where $\hat{G}_n,\hat{G}_m\in\mathfrak{g}$.\footnote{Note our abuse of notation: $\hat{G}_{n\in\mathcal{I}}$ is an element in Hamiltonian only, while $\hat{G}_{n\in\mathcal{J}}$ is an element in the whole Lie algebra $\mathfrak{g}$, for which $\{\hat{G}_n\}|_{n\in\mathcal{I}}\subseteq\mathfrak{g}$.}

When a closed Lie algebra $\mathfrak{g}$ exists, it has been shown that the decomposition (\ref{Definition:one}) exists. The decomposition might not be global, in the sense that it is valid at all times $t$, but it can be produced at least in the vicinity of the origin \cite{Wei:Norman:1963,Wei:Norman:1964}.

\section{Factroizing time evolution into a (finite) sequence of elementary steps}
In this section we proceed to introduce a different way of tackling the problem of factorizing the time-evolution operator than the one outlined above.

\subsection{Ordered factorized time-evolution operators}
The factorization (\ref{time:evolution:operator:decoupled:old}) is obtained as a product of individual elementary terms that each depend on the initial and final time. The advantage of this factorization is that the factorized form is \textit{equivalent} to the original one \textit{for all times} within an appropriate time interval starting at $t_0$, see \cite{Wei:Norman:1964}. However, the dependence on the final time $t$ of each term in the sequence of operations raises the question of the possibility of concretely implementing such operations in realistic setups. In fact, once operation $p$ is done, one would have to subsequently apply operation $p+1$ from time $t_0$ to time $t$ again.

\begin{figure}[ht!]
\includegraphics[width=1\linewidth]{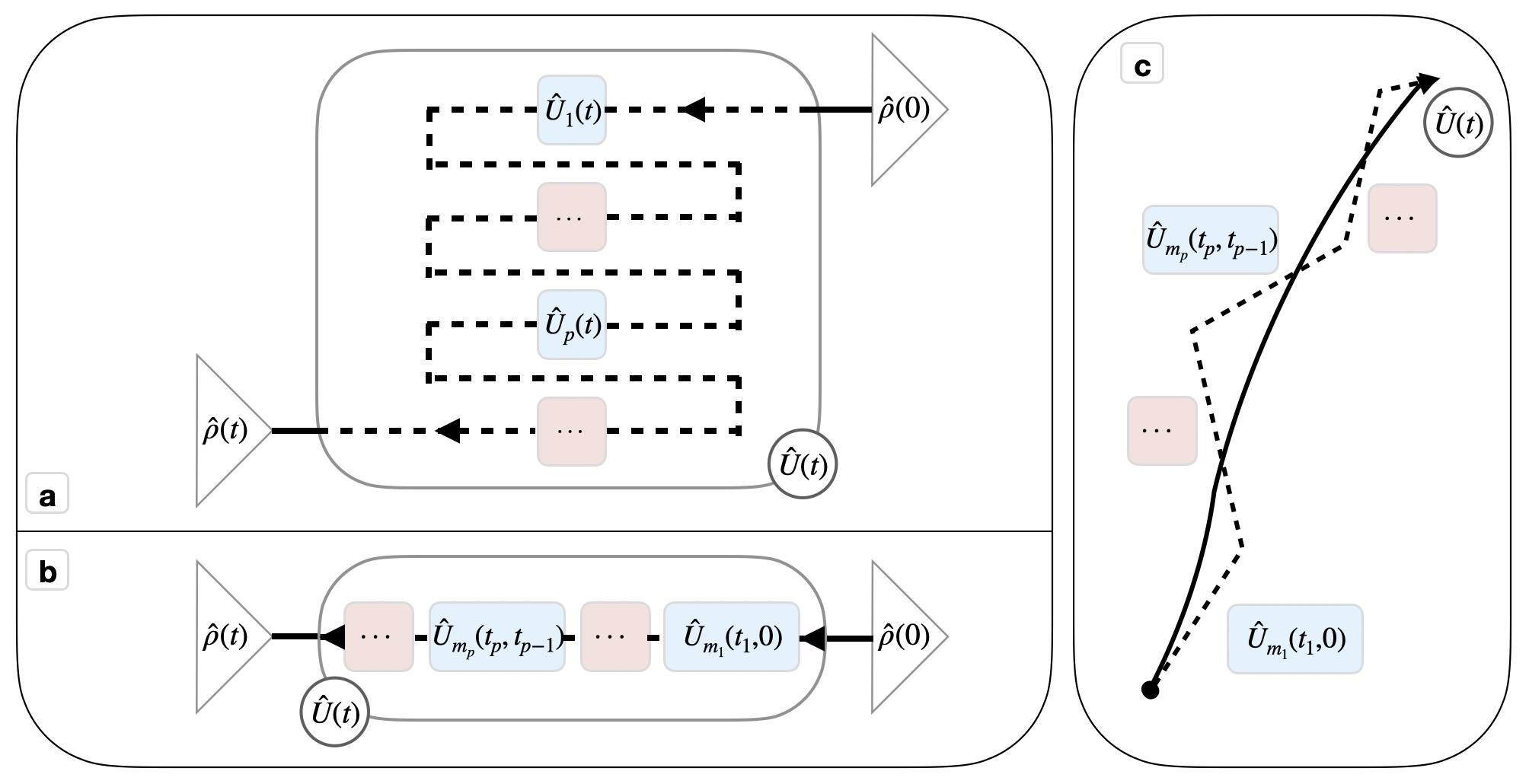}
\caption{The figure represents the two different approaches to factorizing the time-evolution operator: (a) as a sequence of operations that all depend on the initial and final time; (b) as a time-ordered sequence of operations. In both cases, the factorization of the time-evolution operator requires us to apply a set of elementary operations \textit{sequentially}. However, in the former case (a) each operation depends on the full time evolution from $t=0$ to $t$, therefore leaving open the question of causality (see dotted lines). On the other hand, in the latter case (b) the sequence is time-ordered and ``causal'' by construction. The fundamental difference between the two factorizations is that, while in case (a) the factorized operator is equivalent to the original one \textit{at all times} (or at least close enough to the initial time $t_0=0$), in case (b) the decomposition can be obtained for each time $t$ \textit{fixed a priori}. In part (c) it is clear that we might reconstruct (or approximate) the complete time evolution (thick line) with a sequence of operations (dotted line) chosen judiciously.} \label{fig1}
\end{figure}

To solve such conundrum, and therefore propose an expression that would better model a realistic sequence of physical operations, let us introduce the following important new concept of factorization.
\begin{definition}[Ordered factorization] \label{Definition:two}
Given a time $t_{\textrm{f}}$, a Hamiltonian $\hat{H}(t)=\sum_{n\in\mathcal{I}}h_n(t)\,\hat{G}_n$, two sets of indices $\mathcal{J}\supseteq\mathcal{I}$ and $\mathcal{K}$, a set of real functions $F_q:=F_q(t_q,t_{q-1})$ with $q\in\mathcal{K}$ and a set of $p_q\in\mathcal{J}$, we say that the time-evolution operator $\hat{U}(t_{\textrm{f}})$ is \emph{ordered factorized} if it can be put in the form
\begin{align}\label{time:evolution:operator:ordered:decoupled}
\hat{U}(t_{\textrm{f}})=\prod_{q\in\mathcal{K}}\exp\left[-i\,F_q(t_q,t_{q-1})\,\hat{G}_{p_q}\right],
\end{align}
for appropriate functions $F_q$, with $F_q|_{t_q=t_{q-1}}=0$ and $t_m\geq t_{m-1}$. Furthermore, if $|\mathcal{K}|=N<\infty$ then $t_N=t_{\textrm{f}}$, while if $|\mathcal{K}|=\infty$ then $\underset{N\rightarrow\infty}{\textrm{lim}}t_N=t_{\textrm{f}}$.
\end{definition}
The difference between the ordered factorized expression (\ref{time:evolution:operator:ordered:decoupled}) and the previous expression (\ref{time:evolution:operator:decoupled:old}) lies in the fact that: (i) we now consider a product of individual terms that might contain exponentials of the same generator, e.g., $\hat{G}_{p_1}=\hat{G}_{p_3}=...$; (ii) the terms in the factorized expression are driven by the functions $F_q(t_q,t_{q-1})$ which are \textit{ordered in time}; (iii) the expression can be obtained only once a particular final time $t_{\textrm{f}}$ has been chosen. If we wish to consider a different time $t_{\textrm{f}}'$, the expression needs to be evaluated again as explained in the following. In this sense, the ordered factorization (\ref{time:evolution:operator:ordered:decoupled}) can be seen as a formal description of a realistic implementation of a sequence of physical operations. For a pictorial difference between the two factorizations see Figure~\ref{fig1}.

\subsection{Tackling the ordered factorizing algorithm}
We have introduced a second approach to factroizing time-evolution operators. Our ambition is now to provide different algorithms to obtain explicit solutions.
We recall that, in principle, there is no guarantee that such expression (\ref{time:evolution:operator:ordered:decoupled}) exists. Furthermore, this expression has to be determined \textit{ex novo} for each choice of final time $t_{\textrm{f}}$. We now outline a straightforward algorithm to solve this problem.
\begin{algorithm}\label{the:algorithm}
The initial time $t_0\equiv0$, the final time $t_{\textrm{f}}$, and an appropriate operator-norm $||\cdot||$ are initially chosen and fixed. Next, one defines the quantities:
\begin{align*}
	R_m:=&\left|\left|\hat{U}(t)-e^{-i\,F(\tilde{t},t_{m-1})\,\hat{G}_n}\,\prod_{k=1}^{m-1}e^{-i\,F_{k}\,\hat{G}_{q_{k}}}\right|\right|\nonumber\\
	\Delta_m:=&\underset{\tilde{t},F(\tilde{t},t_{m-1}),\hat{G}_n}{\emph{max}}R_m\nonumber\\
	X_m:=&\underset{t,F(\tilde{t},\tilde{t}_{m-1}),\hat{G}_n}{\emph{argmax}}R_m
\end{align*}
together with the definition $R_1:=||\hat{U}(t)-\exp[-i\,F(\tilde{t},0)\,\hat{G}_n]||$ and the short hand form $F_q:=F_q(t_q,t_{q-1})$.
The following steps are then performed:
\begin{itemize}
	\item[a)] Initialization of the algorithm by computing the $\Delta_1$ and the set $X_1\equiv(t_1,F_1,\hat{G}_{q_1})$;
	\item[b)] At step $m$, we compute the quantity $\Delta_m$ and the set $X_m:=(t_m,F_m,\hat{G}_{q_m})$;
\end{itemize}
The protocol ends when $\Delta_p\leq\Delta_{p-1}$ for some finite index $p$.
\end{algorithm}
The algorithm outlined above, in brief, requires maximizing the quantity $R_p$ given the results $R_{k}$ for all $k<p$. It has the following outcomes: 
\begin{itemize}
	\item[i)] success with a finite number $N$ of steps, where $t_N=t_{\textrm{f}}$. Then, $\Delta_N=1$.
	\item[ii)] halt at time $t_{\textrm{h}}=t_N<t_{\textrm{f}}$ after a finite number $N$ of steps. Then, $\Delta_N<1$.
	\item[iii)] a succession of infinite steps, i.e., $\underset{N\rightarrow\infty}{\textrm{lim}}t_N=t_{\textrm{f}}$. Here $\underset{N\rightarrow\infty}{\textrm{lim}}\Delta_N\leq1$.
\end{itemize}
We say that it is \textit{successful} if one can obtain any of these outcomes with one of the ``time slices'' $t_m-t_{m-1}$ being finite, and $\,F(t_m,t_{m-1})$ is not constant.

\subsection{Considerations on the algorithm}
The Algorithm~\ref{the:algorithm} introduced to address the problem of ordered factorization of time-evolution operators is our main proposal. Regardless of its general definition and applicability, it is evident that it might be difficult (if not impossible) to implement in practice. This might be seen as a parallel to a few known problems: first, to those typically encountered in quantum information theory when general measures of entanglement are defined. In fact, many measures have appealing general definitions and are often impossible to compute in practice \cite{Plenio:Virmani:2007}. Entanglement witnesses, for example, provide bounds that can be reliably computed and can be sufficient in many cases for specific given tasks \cite{Terhal:2002,Chruscinski:Sarbicki:2014}. Second, to the problem of obtaining an \textit{exact} reconstruction of a given unitary operator induced by a (usually complicated) Hamiltonian. In this sense, many proposals give techniques that require large  -- or potentially infinite -- numbers of elementary steps that approximate the result up to a desired correction \cite{Leforestier:Bisseling:1991,Kosloff:1994,Tannor:2008,Schaefer:Tal-Ezer:2017}. Below we discuss briefly the possibility of tackling this problem in concrete scenarios

We conclude this part by noting that in the ordered decomposition we expect that the sequence $\hat{G}_{p_q}$, with $q\in\mathcal{K}$, of generators that appears in the exponentials can contain many more elements than the Lie algebra $\mathfrak{g}$ of the Hamitlonian (when such Lie algebra is finite), and therefore contain repeated operators. This is in contrast to the ``standard'' factorization, where the Lie algebra $\mathfrak{g}$ of the Hamiltonian, when closed, allows us to obtain a sequence of exponential terms each induced by a \textit{different} generator $\hat{G}_n$. Finally, optimization over all real functions $F_q=F_q(t_q,t_{q-1})$ is a very burdensome requirement. Therefore, a ``lighter'' version of the algorithm can lift the request of such optimization over all $F_q$ functions by, instead, fixing them to be given functions of time (e.g., $F_q=\alpha\,(t_q-t_{q-1})^n$ or $F_q=\alpha\,\cos(t_q-t_{q-1})$), which have perhaps been shown independently to have the potential of providing a positive outcome.

\subsection{Tackling ordered factorization with realistic schemes}
Obtaining exact solutions to problems is desirable, yet often impossible. In addition, in most realistic situations approximating well enough the solution is sufficient, and approaches in this direction have been developed \cite{Leforestier:Bisseling:1991,Kosloff:1994,Tannor:2008,Schaefer:Tal-Ezer:2017}. Finding the exact action of the operator $\hat{U}(t)$ on arbitrary quantum states would be a great achievement, however, in most cases this is simply not possible.
Factorizing the full time evolution operator into a (possibly finite) sequence of elementary steps can solve the problem of applying a unitary operator to a state, however, it opens up a problem of determining exactly the sequence to be applied, with all of its encoded parameters.
Therefore, we might proceed by relaxing one (or both) of the following: (i) the optimization over all generators $\hat{G}_n$; or (ii) the need to obtain the operator $\hat{U}(t)$ exactly, without margin of error. In practice, this means that a concrete implementation of the time-evolution operator $\hat{U}(t)$ might be constructed by a predetermined sequence of operations (thus relaxing condition (i)), thereby providing an effective time-evolution that is only a (perhaps good) approximation of $\hat{U}(t)$ (thus relaxing condition (ii)).

We emphasize that the most important question to be tackled when studying the viability of the proposed algorithm is: \textit{do Hamiltonians exist such that the factorization of the time-evolution operator includes \emph{at least} one \emph{finite} time slice $\delta t_q:=t_q-t_{q-1}$, and the function $F_q(t_q,t_{q-1})$ is not constant}? If this can be answered in the positive, the proposed algorithm would provide a useful valuable addition to the existing techniques available to tackle factorization of time evolution. 
 
\section{Conclusion}
We have approched the problem of expressing the time evolution of a quantum system as a sequence of elementary operations. Assuming that the elementary operations allowed are drawn from the elements of the Hamiltonian Lie algebra, we have proposed an algorithm to obtain an optimal sequence of such operations that can potentially  reproduce the time-evolution operator exactly. The operation to be performed requires comparing a product of operations that increases by one each step of the algorithm with the desired operator, and optimizing over the operation added in the current step. Success is not guaranteed a priori, which leaves open the task of finding conditions for the convergence of the result. Furthermore, the error in approximating the target operator, as well as the possibility of obtaining concrete solutions, need to be investigated. Finally, proof that at least one elementary operation is induced for a \textit{finite} amount of time would give the proposed algorithm a new advantage with respect to existing solutions.
 We leave it to future work to tackle all of these important and fascinating aspects.

\section*{Acknowledgments}
We thank Shai Machnes, Rosario Vunc Vendiciano, Frank K. Wilhelm and Leila Khouri for reading the manuscript, as well as useful comments and discussions.

\newpage

 \bibliographystyle{splncs04}
 \bibliography{mybibliography}

\end{document}